\documentclass[useAMS,usenatbib]{mn2e}
\usepackage{graphicx}
\usepackage{aas_macros}

\title[Constraints on the space density of CVs]{The space density of cataclysmic variables: constraints from the \emph{ROSAT} North Ecliptic Pole Survey}
\author[M.L. Pretorius et al.]{M. L. Pretorius,$^{1}$\thanks{E-mail: mlp@astro.soton.ac.uk} C. Knigge,$^{1}$ D. O'Donoghue,$^{2}$ J. P. Henry,$^{3}$ I. M. Gioia$^{4}$ 
\newauthor
and C. R. Mullis$^{5}$\\ 
$^{1}$School of Physics and Astronomy, University of Southampton, Highfield, Southampton SO17 1BJ, United Kingdom\\
$^{2}$South African Astronomical Observatory, PO Box 9, Observatory 7935, Cape Town, South Africa\\
$^{3}$Institute for Astronomy, University of Hawai'i, 2680 Woodlawn Drive, Honolulu, HI 96822\\
$^{4}$INAF -- Istituto di Radioastronomia del CNR, Via Gobetti 101, I-40129 Bologna, Italy\\
$^{5}$Department of Astronomy, University of Michigan, 918 Dennison, 500 Church Street, Ann Arbor, MI 48109-1042}

\begin{document}


\pagerange{\pageref{firstpage}--\pageref{lastpage}} \pubyear{}

\maketitle

\label{firstpage}

\begin{abstract}
We use the \emph{ROSAT} North Ecliptic Pole (NEP) survey to construct a small, but purely X-ray flux-limited sample of cataclysmic variable stars (CVs).  The sample includes only 4 systems, 2 of which (RX J1715.6+6856 and RX J1831.7+6511) are new discoveries.  We present time-resolved spectroscopy of the new CVs and measure orbital periods of $1.64\pm0.02\,\mathrm{h}$ and $4.01\pm0.03\,\mathrm{h}$ for RX J1715.6+6856 and RX J1831.7+6511, respectively.  We also estimate distances for all the CVs in our sample, based mainly on their apparent brightness in the infrared.  The space density of the CV population represented by our small sample is $1.1^{+2.3}_{-0.7} \times 10^{-5}\,\mathrm{pc^{-3}}$.  We can also place upper limits on the space density of any sub-population of CVs too faint to be included in the NEP survey.  In particular, we show that if the overall space density of CVs is as high as $2 \times 10^{-4}\,\mathrm{pc^{-3}}$  (as has been predicted theoretically), the vast majority of CVs must be fainter than $L_X \simeq 2 \times 10^{29}\,\mathrm{erg\,s^{-1}}$.

\end{abstract}


\section{Introduction}
Cataclysmic variable stars (CVs) are semi-detached interacting binary stars, consisting of a white dwarf primary accreting from a companion that is usually a late-type, approximately main-sequence star.  \citet{bible} gives a comprehensive review of the subject.

Mass transfer, and the evolution of CVs, is driven by angular momentum loss from the orbit.  At long orbital periods ($P_{orb}\ga3\,\mathrm{h}$), angular momentum loss through magnetic braking is thought to dominate over that resulting from gravitational radiation.  The canonical disrupted magnetic braking model of CV evolution was designed to explain the period gap (a pronounced drop in the number of CVs at $2~\mathrm{h} \la P_{orb}\la3$~h).  In this scenario, magnetic braking stops when $P_{orb}$ reaches $\simeq 3\,\mathrm{h}$, the secondary loses contact with its Roche lobe, and gravitational radiation is left as the only angular momentum loss mechanism (e.g. \citealt{RobinsonBarkerCochran81}; \citealt{RappaportVerbuntJoss83}; \citealt{SpruitRitter83}).  The secondary regains contact with its Roche lobe when gravitational radiation has decreased $P_{orb}$ to $\simeq 2\,\mathrm{h}$, and mass transfer resumes.  As the secondary star loses mass, its thermal time-scale eventually exceeds the mass-transfer time-scale (even though the mass-loss time-scale increases as $M_2$ decreases).  When this happens, the secondary is not able to shrink rapidly enough in response to mass loss, so that the orbital evolution slowly moves back through longer periods (e.g. \citealt{Paczynski81}; \citealt{PaczynskiSienkiewicz81}; \citealt{RappaportJossWebbink82}).  CVs in this final phase of evolution, where $P_{orb}$ is increasing, are referred to as `period bouncers'.  This reversal in the direction of change in $P_{orb}$ causes a sharp cut-off in the $P_{orb}$ distribution of hydrogen-rich CVs at about 76~min, called the period minimum.

Several serious discrepancies between the predictions of theory and the properties of the observed CV sample remain.  Theoretical models persistently place the period minimum about 10~min short of the observed cut-off at $P_{orb}\simeq 76\,\mathrm{min}$ (see e.g. \citealt{Kolb02} for a review).  It has also long been suspected that the predicted ratio of short- to long-period CVs is larger than is implied by observations (e.g. \citealt{Patterson98}).  \cite{PretoriusKniggeKolb07} recently confirmed that this is a real discrepancy, rather than the result of observational bias.  

In order to make progress, it is necessary to place quantitative constraints on other key parameters predicted by CV formation and evolution scenarios. The most fundamental of these is the overall space density of CVs, $\rho$. However, observational estimates of $\rho$ differ by more than an order of magnitude at present, and little effort has been made to quantify the statistical and systematic uncertainties affecting these estimates. In order to understand this situation, it is useful to take a closer look at some of the more extreme values in the literature.

\cite{CieslinskiDiazMennickent03} suggested a very low space density of $\rho \le 5 \times 10^{-7}\mathrm{pc}^{-3}$ for dwarf novae (DNe), the most common type of CV. This result was based on DN outbursts detected in the OGLE-II survey, and assumes that most DNe in the OGLE-II fields within 1 kpc would have been recovered. However, the time coverage of the survey ($\simeq 4$~y) was much shorter than the longest DN recurrence time-scales (e.g. $\simeq 30$~y for WZ Sge), and long inter-outburst intervals are known to be associated precisely with the intrinsically faintest CVs that probably dominate the overall population. At the other end of the spectrum, \cite{SharaMoffatPotter93} obtained a rather high estimate of $\rho \sim  10^{-4}\mathrm{pc}^{-3}$ from a photometric survey for faint (possibly hibernating) CVs, but here the CV nature of the objects that were counted remains to be established. Finally, \cite{SchwopeBrunnerBuckley02} obtained $\sim 3 \times 10^{-5}\,\mathrm{pc^{-3}}$ from a CV sample constructed from the \emph{ROSAT} Bright Survey. However, this number was primarily based on two systems with estimated distances of $\sim 30$~pc, both of which were subsequently shown to be 5--10 times more distant \citep{ThorstensenLepineShara06}.

These examples illustrate that the statistical errors associated with observational space density estimates are usually dominated by uncertain distances and small number statistics, whereas the dominant systematic errors are caused by selection effects. We also suspect that the hard-to-quantify systematic uncertainties usually outweigh the statistical errors, which probably explains why virtually no space density estimate in the literature has come with an error bar. Given all this, our goal here is to provide a new determination of $\rho$ from a small CV sample with very simple, well-defined selection criteria. Since our sample does not suffer from unquantifiable systematic errors, we are also able to provide meaningful errors on $\rho$ that fully account for small number statistics and uncertain distances.

An attribute that should be shared by all CVs is X-ray emission generated in the accretion flow.  This is useful, because it is much easier to construct a reasonably deep flux-limited CV sample in X-rays than it is in the optical.  For an optical sample, it is usually necessary to cut down on the number of sources that can be followed up by introducing additional selection criteria based on, e.g., colour, variability, or emission lines; any such selection makes it more difficult to understand the completeness of the resulting sample.  Another advantage of an X-ray flux-limited sample is that $F_X/F_{opt}$ decreases with increasing $P_{orb}$ (i.e. with increasing optical luminosity).  This means that CVs do not span as wide a range in X-ray as in optical luminosity.

The \emph{ROSAT} North Ecliptic Pole (NEP) survey (e.g. \citealt{Gioia03}; \citealt{Henry06}) is constructed using \emph{ROSAT} All Sky Survey (RASS; see \citealt{rosatbsc} and \citealt{rosatfsc}) data around the north ecliptic pole; the RASS reaches its greatest sensitivity in this 81~sq.deg. region, centred on $\alpha_{2000} = 18^{\mathrm{h}}00^{\mathrm{m}}$, $\delta_{2000} = +66^\circ 33'$.  The NEP survey thus comprises the deepest wide-angle, contiguous region ever observed in X-rays; it consists of a complete sample of 442 sources above a flux of about $10^{-14}\mathrm{erg\,cm^{-2}s^{-1}}$ in the $0.5$ to $2.0$~keV band (the flux limit varies from $1.2 \times 10^{-14}$ to $9.5 \times 10^{-14}\mathrm{erg\,cm^{-2}s^{-1}}$ over the survey area\footnote{Assuming a 10~keV thermal bremsstrahlung spectrum; this choice is justified in Section~\ref{sec:rho}.}).  This unique combination of depth and breadth, together with moderate Galactic latitude and extinction, provides an excellent opportunity to investigate the space density of CVs.  Very importantly, out of the 442 X-ray sources included in the survey, all but 2 (the number of spurious sources expected in a sample like this; \citealt{Gioia03}) have been optically identified and spectroscopically observed.  Therefore, we are able to construct a complete, purely X-ray flux-limited CV sample from the NEP survey.

We use the CVs detected in the \emph{ROSAT} NEP survey to provide robust observational constraints on the CV space density, by carefully considering the uncertainties involved.  The constraints include an estimate based on the observed sample of CVs, as well as upper limits on the size of a hypothetical population of CVs with X-ray luminosities low enough to evade the survey entirely.  We also present radial velocity studies of the two new CVs discovered in this survey.

\section{CVs in the \emph{ROSAT} NEP survey}
The NEP survey area includes 3 known CVs, namely IX Dra, SDSS J173008.38+624754.7 (we will abbreviate this to SDSS J1730), and EX Dra.  All three were detected in the RASS, but IX Dra is slightly too faint to be formally included in the NEP sample.

We inspected the optical spectra of all Galactic NEP sources, and found only two additional objects with the spectral appearance of interacting binaries.  A more detailed study of stellar sources included in the survey has yielded accurate spectral classifications of all stellar NEP sources \citep{MicelaAfferFavata07}, and no additional CV candidates.  We are therefore confident that we have identified every CV in the NEP sample.

\begin{figure}
 $\begin{array}{c@{\hspace{1mm}}c}
 \includegraphics[width=41.5mm]{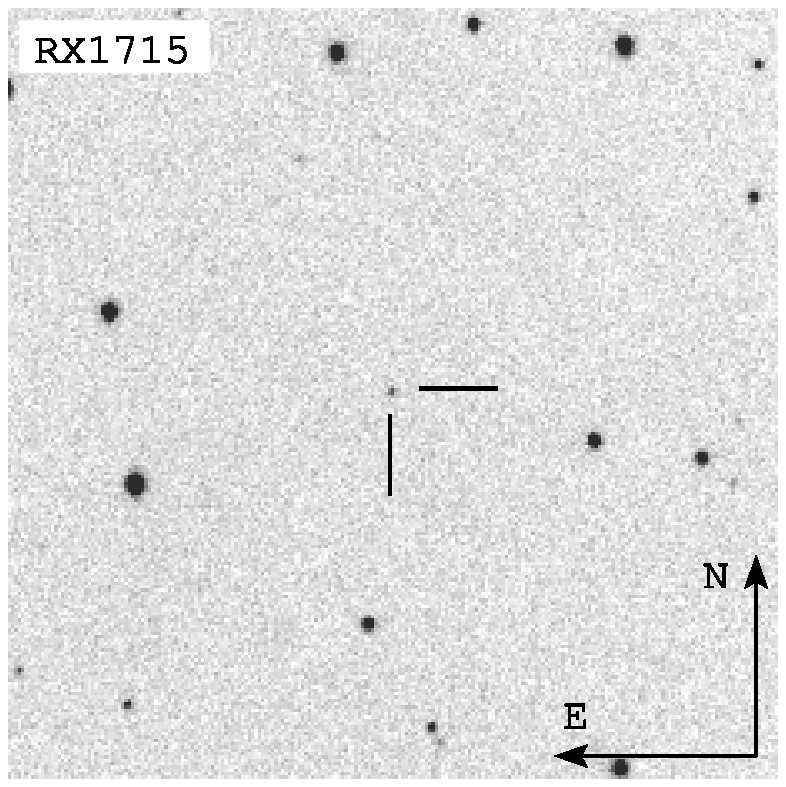} &
 \includegraphics[width=41.5mm]{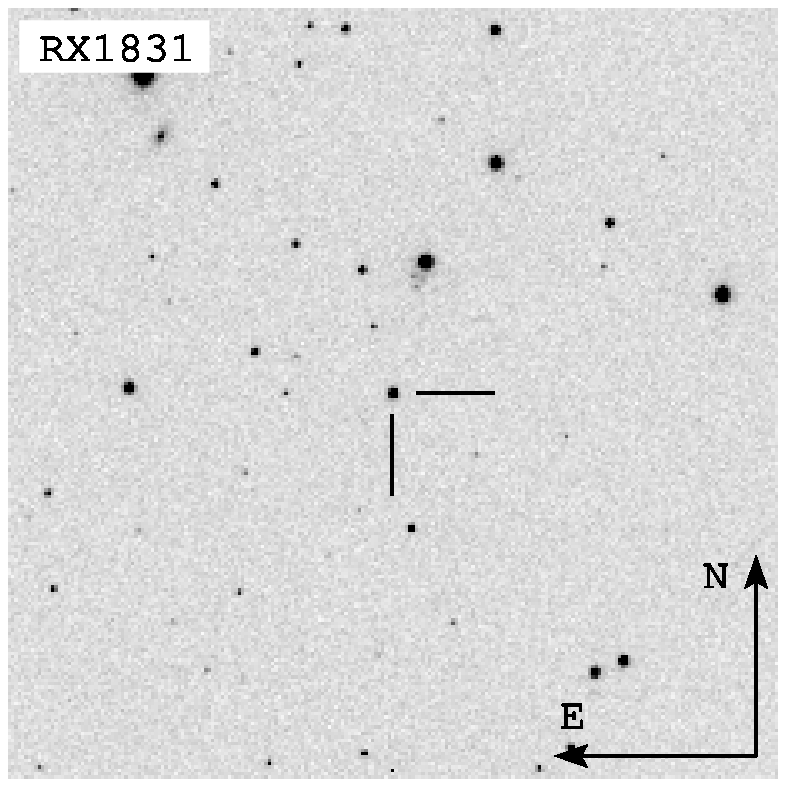} 
 \end{array}$
  \caption {$4' \times 4'$ finding charts of the optical counterparts of RX J1715.6+6856 (left) and RX J1831.7+6511 (right), made using white light acquisition images taken with CAFOS.  North is at the top and east to the left in both images.}
 \label{fig:find}
\end{figure}

The two interacting binary candidates are RX J1715.6+6856 and RX J1831.7+6511, which were confirmed as CVs by the follow-up observations described below.  Fig.~\ref{fig:find} gives finding charts for the two newly discovered CVs; their coordinates are listed in Table~\ref{tab:log}.  \cite{Gioia03} report photographic magnitudes of $E=18.6$ and $O=18.3$ for RX J1715.6+6856, and $E=14.8$ and $O=15.2$ for RX J1831.7+6511 (the errors in these measurements are $\pm0.5$~mag; $O$ and $E$ are similar to Johnson $B$ and Kron-Cousins $R$, respectively).

\subsection{Observations of the newly discovered CVs}
\label{sec:obs}
In order to confirm that RX J1715.6+6856 and RX J1831.7+6511 are CVs, and to measure their orbital periods, we obtained medium-resolution, time-resolved spectroscopy with the Calar Alto Faint Object Spectrograph (CAFOS) on the Calar Alto 2.2-m telescope.  Table~\ref{tab:log} gives a log of the observations.  The G-100 grism was used in combination with a slit width of $1\farcs2$, yielding a spectral resolution of $\simeq 4.2\,\mathrm{\AA}$\ over the wavelength range 4\,300 to 8\,000~\AA.  A few spectra of each object were also taken using the B-200 grism, and a slit width of $1\farcs5$.  These spectra have lower resolution ($\simeq 9.4\,\mathrm{\AA}$) and were not used for radial velocity measurements.  The spectrophotometric standard star BD+28$^\circ$4211 \citep{oke90} was observed on all three nights to provide flux calibration.  Except for the first hour of the first night, conditions were photometric during our observations.  Regular arc lamp exposures were taken to maintain an accurate wavelength calibration.

\begin{table*}
 \centering
  \caption{Log of the observations, and coordinates for the optical counterparts of the two X-ray sources.}
  \label{tab:log}
  \begin{tabular}{@{}llllllll@{}}
  \hline
Object        & $\alpha_{2000}$ & $\delta_{2000}$ & Date at the start & First integration  & Grism & integration     & Number of \\
              &                 &                 & of the night      & HJD $2453900.0+$   &       &  time/s         & spectra   \\
 \hline
RX J1715.6+6856 & 17:15:41.7      & +68:56:43       & 2006 Jul 27       & 44.36780950        & B-200 & 1\,000      & 7    \\
       &                 &                 & 2006 Jul 28       & 45.37354850        & G-100 & 1\,000      & 14   \\
       &                 &                 & 2006 Jul 29       & 46.35562975        & G-100 &    800      & 9   \\[0.1cm]
RX J1831.7+6511 & 18:31:44.4      & +65:11:32       & 2006 Jul 27       & 44.51118822        & G-100 &    600      & 16   \\
       &                 &                 &                   & 44.66093472        & B-200 &    600      & 2    \\
       &                 &                 & 2006 Jul 28       & 45.59054418        & G-100 &    600      & 10   \\
       &                 &                 & 2006 Jul 29       & 46.47792121        & G-100 &    600      & 22   \\
 \hline
 \end{tabular}
\end{table*}

The data were reduced using standard procedures in IRAF\footnote{IRAF is distributed by the National Optical Observatories.}, including optimal extraction \citep{Horne86}.  We computed radial velocities of both the H$\alpha$ and H$\beta$ lines, using the Fourier cross correlation method described by \cite{fxcor}, as implemented in the fxcor routine in IRAF.  In the case of the H$\alpha$ line, we correlated the wavelength range 6\,450 to 6\,650~\AA, while for the measurements of H$\beta$ radial velocities, the correlation was restricted to the range from 4\,700 to 5\,020~\AA.  For both RX J1715.6+6856 and RX J1831.7+6511, the H$\alpha$ line yielded higher $S/N$  radial velocity curves; we therefore use the radial velocities measured for this line in determining the orbital periods.

\subsubsection{Observational results for RX J1715.6+6856}
\begin{figure}
 \includegraphics[width=84mm]{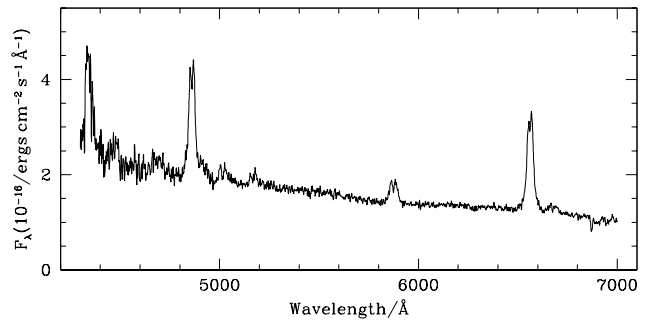}
  \caption {The average of all the spectra of RX J1715.6+6856 taken with the G-100 grating.  Individual spectra were shifted to the rest frame using the radial velocities calculated from the H$\alpha$ line before being averaged.  The system is identified as a CV by its broad, double peaked Balmer, He\,{\scriptsize I}, and He\,{\scriptsize II} emission lines.  Also present is the Fe\,{\scriptsize II}\,$\lambda$5169 line.
}
 \label{fig:rx1715spec}
\end{figure}

Fig.~\ref{fig:rx1715spec} shows the average spectrum of RX J1715.6+6856, which displays broad, double peaked Balmer and He\,{\scriptsize I} emission lines superimposed on a blue continuum; He\,{\scriptsize II}\,$\lambda$4686 and Fe\,{\scriptsize II}\,$\lambda$5169 are also detected in emission.  The system has the spectral appearance of a quiescent DN.  No large amplitude brightness variations were evident in our data, and we estimate $V=18.3 \pm 0.2$ from our spectroscopy (where the uncertainty is dominated by slit losses).  

\begin{figure}
 \includegraphics[width=84mm]{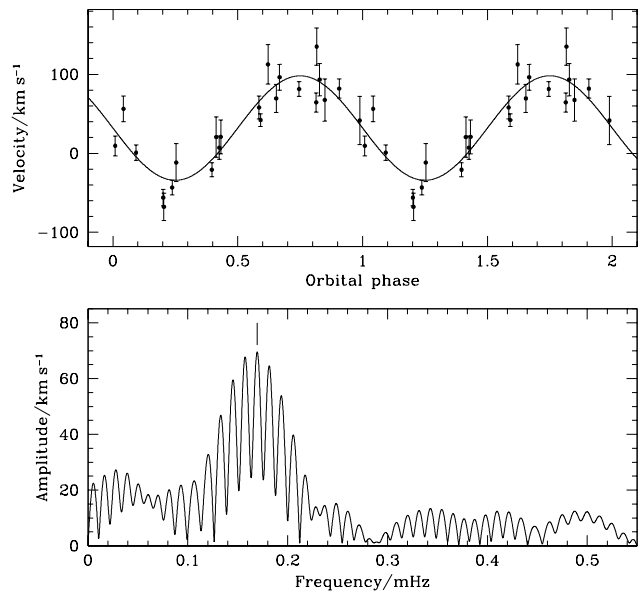} 
  \caption {Lower panel: The Fourier transform of the H$\alpha$ radial velocity curve of RX J1715.6+6856 obtained on 28 and 29 July 2006.  A vertical bar at 0.1695~mHz (1.639~h) indicates the period that provides the best fit to our data.  Upper panel: The radial velocity measurements, folded in the best-fit period.  One cycle is repeated in this plot; the superimposed sine function is a non-linear least squares fit to the data.
}
 \label{fig:rx1715ft}
\end{figure}

At $V = 18.3$, RX J1715.6+6856 is close to the magnitude limit of our telescope/instrument combination, and the time resolution and base line of our data yield only a fairly uncertain period.  The Fourier transform in Fig.~\ref{fig:rx1715ft} shows that the radial velocity curve is aliased; the largest amplitude signal is at 0.1695~mHz (corresponding to a period of 1.639~h), and is marked by a vertical bar in the lower panel of Fig.~\ref{fig:rx1715ft}.  The two nearest peaks are at 0.1575 and 0.1816~mHz.  We fitted sinusoids to the data by non-linear least squares, using the periods and amplitudes of these three signals as starting values.  We found that the 1.639~h period provided a significantly better fit than the other aliases (the standard deviation of the residuals for a fit near this period is $\simeq15$ \% lower than for either of the other choices).  The 1.639~h period also yields the largest best-fit amplitude (66.1~km/s).  Nevertheless, there is some uncertainty in the cycle count.  Assuming that we have identified the correct alias, the orbital period is of RX J1715.6+6856 is $1.64\pm0.02\,\mathrm{h}$.  The top panel of  Fig.~\ref{fig:rx1715ft} shows the radial velocities of H$\alpha$, phase-folded using this period, with the sine fit over-plotted.

\subsubsection{Observational results for RX J1831.7+6511}
\begin{figure}
 \includegraphics[width=84mm]{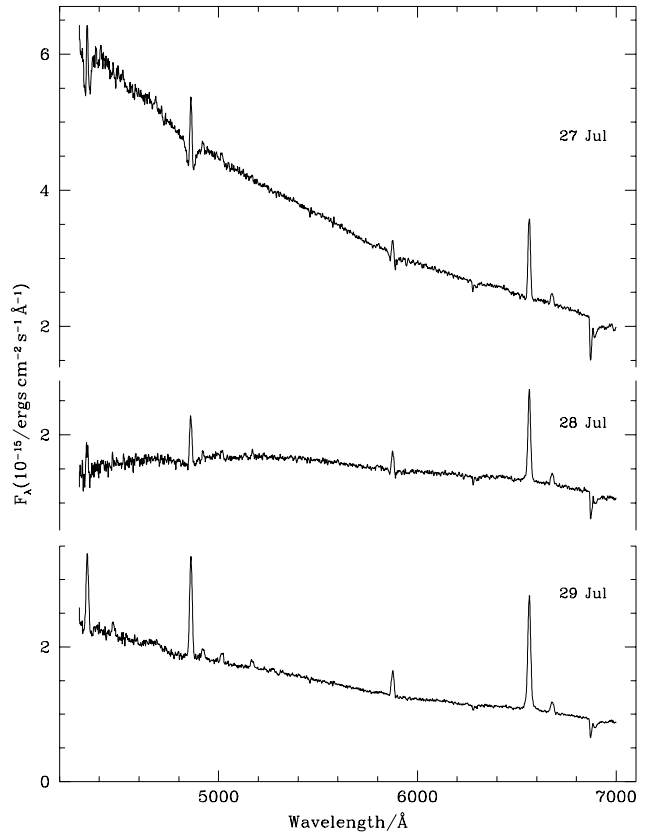}
 \caption {The average nightly spectrum of RX J1831.7+6511.  The system faded from $V=15.0$ to $V=15.9$ over the three nights, with the strength of the emission lines increasing, as is commonly seen during the decline from dwarf nova outbursts (e.g. \citealt{HessmanRobinsonNather84}).
}
 \label{fig:rx1831spec}
\end{figure}

RX J1831.7+6511 was descending into a photometric low state over the three nights of our observations, fading by roughly a magnitude in $V$.  This was most likely the decline from a dwarf nova outburst.
The spectra show fairly narrow Balmer and He\,{\scriptsize I} emission lines (the $FWHM$ of the H$\alpha$ line is $\simeq800\,\mathrm{km/s}$).  The H$\beta$ and H$\gamma$ line profiles change from broad absorption wings with emission cores in the bright state, to pure emission in the faint state.  This transition is illustrated in Fig.~\ref{fig:rx1831spec}, which displays nightly averages of the G-100 spectra after shifting to the rest frame using the radial velocities calculated from the H$\alpha$ line.  We also detect Fe\,{\scriptsize II}\,$\lambda$5169 in RX J1831.7+6511, increasing in strength as the system fades.

\begin{figure}
 \includegraphics[width=84mm]{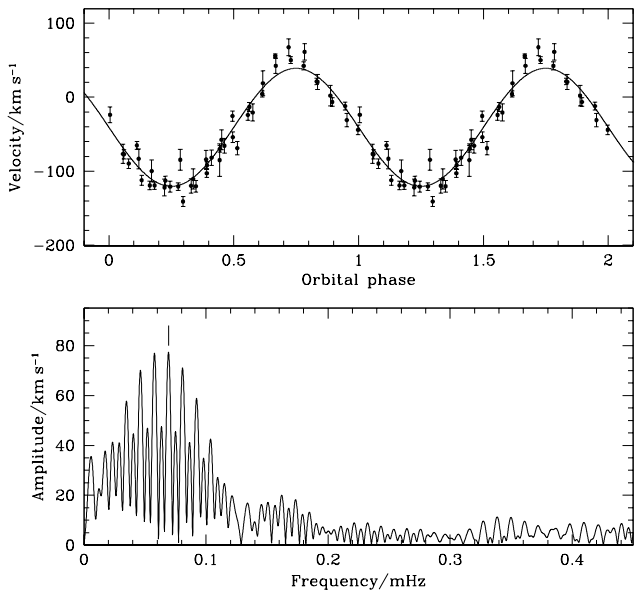} 
  \caption {The low frequency end of the Fourier transform of the H$\alpha$ radial velocity curve of RX J1831.7+6511 (bottom panel).  A vertical bar at 0.0693~mHz (4.01~h) marks the signal that most likely corresponds to the orbital period.  The top panel shows the phase-folded radial velocity curve, with the best-fit sine function over-plotted (all data are plotted twice).
}
 \label{fig:rx1831ft}
\end{figure}

The lower panel of Fig.~\ref{fig:rx1831ft} shows a Fourier transform of the radial velocity curve of RX J1831.7+6511 (including all 3 nights of data).  A vertical bar at 0.0693~mHz (4.01~h) indicates the highest amplitude signal, but there is also strong power present at the one-day aliases (0.0578~mHz and 0.0808~mHz).  Non-linear least-squares fits of sine functions to the data produce the largest amplitude for the 4.01~h period; the standard deviation of the residuals is also roughly 45\% larger for fits using the other two candidate periods.  We are therefore reasonably confident that the signal at 4.01~h represents the orbital modulation, and find $P_{orb}=4.01\pm0.03\,\mathrm{h}$.  The top panel of Fig.~\ref{fig:rx1831ft} shows the H$\alpha$ radial velocity curve, phase-folded using the orbital period, together with the best-fit sine function.  The ampliude of the fit is 79.7~km/s.

\subsection{Distance estimates}
\label{sec:dist}
Several independent methods of estimating distances to CVs are discussed by \cite{Patterson84} and \cite{brian87}.  We will use methods based on DN outburst maximum, the strength of emission lines, and the brightness of the secondary star.

The relation between orbital period and the absolute $V$ magnitudes of DNe at maximum found by \cite{brian87}, as recalibrated by \cite{HarrisonJohnsonMcArthur04}, is
\begin{equation}
M_{Vmax}=5.92-0.383P_{orb}/\mathrm{h}.
\end{equation}
In this equation, $M_{Vmax}$ is corrected for binary inclination ($i$) using
\begin{equation}
\Delta M_V(i)=-2.50 \log \left[ \left(1+1.5\cos i \right)\cos i \right]
\label{eq:icorr}
\end{equation}
\citep{PaczynskiSchwarzenberg-Czerny80}.  To apply this relation one thus needs knowledge of $i$, as well as a measurement of $V$ at maximum.  The scatter of data around the $M_{Vmax}$--$P_{orb}$ relation is roughly $\pm0.5$~mag.  

The equivalent widths ($EW$; throughout this paper we will take the $EW$s of emission lines as positive) of disc emission lines are theoretically expected to increase with decreasing mass transfer rate ($\dot{M}$); 
\cite{Patterson84} gives an empirical relation 
\begin{equation}
\mathrm{EW(H}\beta)=0.3M_{V,disc}^2+\mathrm{e}^{0.55(M_V-4)}
\end{equation}
that predicts $M_{V,disc}$ (the absolute $V$ magnitude of the disc) to within about 1.5~mag for $EW(H\beta)\ga 15\,\mathrm{\AA}$.  For EW(H$\beta$)$\la 15\,\mathrm{\AA}$, the data imply only $M_{V,disc}\la+6$.

\cite{Bailey81} describes a method of estimating distances to CVs that relies on finding the $K$-band surface brightness of the secondary from $V-K$ colour of the system.  It has since been shown that this method is not as reliable as was first thought, and that the $V$ and $K$ magnitudes of a CV alone cannot provide a secure distance estimate (e.g. \citealt{Beuermann00}).  \cite{Knigge06} provides an alternative, based on a semi-empirical donor sequence for CVs that gives absolute magnitudes for the secondaries as a function of $P_{orb}$.  \cite{Knigge06} finds offsets of $\Delta J = 1.56$, $\Delta H = 1.34$, and $\Delta K = 1.21$ between his sequence and the absolute $JHK$ magnitudes of a sample of CVs with parallax distances.  The scatter of the sample magnitudes around the offset sequence are $\sigma_{\Delta J} = 1.25$, $\sigma_{\Delta H} = 1.12$, and $\sigma_{\Delta K} = 1.07$.  We will use these offsets and errors to obtain distance estimates; the limitations of the method are discussed in more detail by \cite{Knigge06}.

The infrared observations used here are from the Two Micron All Sky Survey (2MASS; \citealt{2mass}).  We interpolate linearly on the high resolution version of the donor sequence to find predicted absolute magnitudes for the secondaries at the orbital periods of the observed CVs.  The CIT magnitudes of \cite{Knigge06} were transformed to the 2MASS system using the prescription of \cite{Carpenter01}, updated to reflect the final 2MASS data release\footnote{The updated colour transformations are available at http://www.astro.caltech.edu/$\sim$jmc/2mass/v3/transformations/.}.  Errors from the transformation are insignificant for our purposes, and will not be quoted (although we did include them).

To estimate extinction effects, we compared a model of \cite{AmoresLepine05}\footnote{The model presented in that paper that takes into account the spiral structure of the Galaxy.} to the model of \cite{DrimmelCabrera-LaversLopez-Corredoira03} at several positions in the NEP region, and found an average difference of $\simeq$45 \% between the $A_V$ values predicted by the two.  We will use the \cite{AmoresLepine05} extinction values, and assume errors of 50 \%.  Extinction is obviously a fuction of distance; we therefore do a few iterations to ensure that the value we finally use in the distance calculation is that given by the model at the estimated distance to the object; for lower limits on distances we use the total Galactic extinction.  To convert from visual extinction to that in the 2MASS bands, we use $A_J = 0.282 A_V$, $A_H=0.175 A_V$, and $A_{K_S} = 0.112 A_V$ \citep{CambresyBeichmanJarrett02}.  

We assume that errors in absolute magnitude, apparent magnitude, and extinction are Gaussian, and use this to find a probability distribution for the distance to each source.  The distances we quote are in all cases the median, together with the symmetric 1-$\sigma$ confidence interval around the median (this is the interval defined so that the integral of the distribution between its lower bound and the median, as well as between the median and its upper bound, is 34\% of the integral over all distances).

Two of the three techniques we use to estimate distances require the orbital period of the CV.  For RX J1715.6+6856, RX J1831.7+6511, and SDSS J1730 there is some cycle count ambiguity in the $P_{orb}$ measurements, and there is also some uncertainty in the orbital period of IX Dra (see below).  However, the period uncertainty is in all cases negligible for the distance estimates given here.

\subsubsection{RX J1715.6+6856}
RX J1715.6+6856 was not detected in 2MASS; we estimate $K_{S}>15.6$ from the magnitude of the faintest stars detected in the field.  At the period of RX J1715.6+6856 (1.64~h), the donor sequence predicts $M_{K_S}=8.74$ for the secondary star (if this system is not a period bouncer; below we make the same assumption for SDSS J1730), giving $d > 235\,\mathrm{pc}$.  

Using the $M_{V,disc}$--$EW(H\beta)$ relation, and $EW(H\beta) = 62\,\mathrm{\AA}$ (from our spectroscopy) we obtain $M_V = 10.2 \pm 1.5$ (assuming that the disc dominates the flux in $V$), and with $A_V=0.08 \pm 0.04$, we have $d = 400^{+400}_{-200}\,\mathrm{pc}$.

\subsubsection{SDSS J1730}
The discovery of SDSS J1730 was announced by \cite{sdsscvs1}, who observed a DN outburst, and measured $EW(H\beta)=63\,\mathrm{\AA}$ (in quiescence) and an orbital period of roughly 1.95~h; the refined value is $P_{orb}=1.837 \pm 0.002\,\mathrm{h}$ (John Thorstensen, private communication).  As expected at this orbital period, SDSS J1730 is an SU UMa star (vsnet-alert 6798); the fractional superhump period excess indicates that it has not evolved to the period minimum \citep{PattersonKempHarvey05}.  

\cite{Knigge06} gives $M_{K_S} = 8.22$ for a typical secondary at this period, and from the observed $K_S = 15.22 \pm 0.18$ we therefore obtain a lower limit of $d > 250\,\mathrm{pc}$, and an estimate of $d=440^{+280}_{-170}\,\mathrm{pc}$ (using $A_V=0.08 \pm 0.04$). 

The observed $EW(H\beta)$ gives $M_{V,disc}=10.3 \pm 1.5$, and with $V=15.9 \pm 0.1$, we have $d = 130^{+130}_{-60}\,\mathrm{pc}$ (again assuming that the $V$-band flux originates only from the disc).  Since this distance is only barely consistent with the robust lower limit, and since any white dwarf contribution to the $V$-band flux will imply that it is too low, we will adopt the donor-based distance estimate.

\subsubsection{RX J1831.7+6511}
RX J1831.7+6511 has $P_{orb}=4.01\,\mathrm{h}$ (implying $M_{K_S} = 6.20$ for the secondary), $K_S = 14.96 \pm 0.14$, and $A_V=0.11 \pm 0.06$ (i.e. $A_{K_S}=0.012 \pm 0.007$).  This gives a lower limit of $d > 559\,\mathrm{pc}$, and an estimate of $d=980^{+630}_{-380}\,\mathrm{pc}$.

For comparison, we measure $EW(H\beta)=14\,\mathrm{\AA}$ and $V=15.9 \pm 0.2$ from our 29 July 2006 spectra.  Assuming the system was in quiescence, the $M_{V,disc}$--$EW(H\beta)$ relation gives $M_V \la +6$, and thus $d \ga 900\,\mathrm{pc}$.

\subsubsection{EX Dra}
Being a deeply eclipsing system, EX Dra has been the subject of several detailed studies.  It is a U Gem star with $P_{orb}=5.0385$~h (\citealt{FiedlerBarwigMantel97}), a secondary spectral type between M1V and M2V (\citealt{BaptistaCatalanCosta00}; \citealt{BillingtonMarshDhillon96}; c.f. \citealt{HarrisonOsborneHowell05}), and an orbital inclination of $i=85^{+3}_{-2}\,\mathrm{deg.}$ (\citealt{BaptistaCatalanCosta00}; \citealt{FiedlerBarwigMantel97} obtained a similar value).  The out-of eclipse quiescent $V$ magnitude is 15.3, and the system reaches $V=12.3$ at maximum.  At mid-eclipse in quiescence, $V=16.66\pm0.07$, $I=14.46 \pm 0.05$, and essentially all of the light originates from the secondary star (\citealt{BaptistaCatalanCosta00}; \citealt{BaptistaCatalan01}; \citealt{ShafterHolland03}).  Mid-eclipse photometry has provided distance estimates based on the photometric parallax of the secondary of $240\pm90\,\mathrm{pc}$ \citep{ShafterHolland03} and $290\pm80\,\mathrm{pc}$ \citep{BaptistaCatalanCosta00}.

As a check, we can also use the CV donor sequence of \cite{Knigge06} to estimate the distance to EX Dra.  We assume here that in quiescence, the secondary is the only source of mid-eclipse light.  The sequence gives $M_V=8.80\pm0.37$ for secondaries with spectral types in the range M1 to M2.  This gives a distance of $360^{+70}_{-60}\,\mathrm{pc}$.  At $P_{orb}=5.0385$~h, predicted magnitudes for the secondary star are $M_V=9.56 \pm 0.96 $ and $M_I = 7.51 \pm 0.77$ (the errors are derived from an error of $\pm1$ spectral type in the sequence, and are probably conservative).  If we use the measured period rather than spectral type, we find distance estimates of $260^{+140}_{-90}$ and $240^{+120}_{-80}\,\mathrm{pc}$ from the apparent mid-eclipse $V$ and $I$ magnitudes, respectively.  The first of these estimates uses $A_V=0.08 \pm 0.04$, while for the other two we set $A_V=0.06 \pm 0.03$.  The reason these distances differ is that EX Dra is not exactly on the donor sequence (see fig.~9 of \citealt{Knigge06}).

The $M_{Vmax}$--$P_{orb}$ relation gives $d = 140^{+60}_{-50}\,\mathrm{pc}$.  A likely reason for this small value is that the dependence on $i$ given by equation~\ref{eq:icorr} is too sensitive at large $i$, because radiation from the disc rim becomes significant \citep{Smak94}.  

Our three donor-based distance estimates and those of \cite{ShafterHolland03} and \cite{BaptistaCatalanCosta00} are, of course, not independent, but they are all formally consistent with each other.  We will adopt a distance of $260^{+140}_{-90}\,\mathrm{pc}$ for this system.

\subsubsection{IX Dra}
\cite{LiuHuZhu99} confirmed IX Dra as a CV.  It is a member of the group of SU UMa stars with unusually short outburst recurrence times, sometimes referred to as ER UMa stars \citep{IshiokaKatoUemura2001}.  The quiescent brightness is $V=17.5$, and maximum of normal outburst is roughly $V=15.4$.
\cite{OlechZloczewskiMularczyk04} measured a superhump period of 1.6072~h, and found a second photometric period of 1.595~h in superoutburst light curves.  If the second period is interpreted as $P_{orb}$, the small fractional superhump period excess ($\epsilon=(P_{sh}-P_{orb})/P_{orb}$) implies a very low ratio of secondary to white dwarf mass \citep{Patterson98}, and thus that IX Dra has a secondary of substellar mass \citep{OlechZloczewskiMularczyk04}.  However, the period bouncer status of IX Dra is suspect for two reasons.  The first is that the high outburst duty cycle (about 30\%) implies a high $\dot{M}$ (however, note that DI UMa has very low $\epsilon$, and similar outburst characteristics; \citealt{Patterson98}).  More importantly, it relies on the 1.595~h signal being the orbital period; a large amplitude (in this case about 0.05 mag) photometric orbital modulation is not usually seen during superoutburst, and the superhump modulation makes it dificult to measure other signals reliably.  Confirmation of the orbital period is needed to decide whether this system is likely to be a period bouncer. 

The inclination of IX Dra is not known, but the fact that there is no eclipse (even at supermaximum) constrains it to $< 70^\circ$.  This gives a limit of $d \ga 545\,\mathrm{pc}$ from the $M_{Vmax}$--$P_{orb}$ relation.  

2MASS provides only a lower limit on $K_S$ and a marginal detection in $H$ for IX Dra; we therefore use $J=16.47 \pm 0.12$.  If we used the period bouncer branch of the donor sequence, we would obtain a distance estimate that is inconsistent with the lower limit given above.  We therefore use the pre-period minimum branch of the donor sequence ($M_J=9.85$), despite the possibility that IX Dra is a period bouncer.  In the direction of IX Dra, $A_V \simeq 0.1$ for distances greater than $\simeq 400\,\mathrm{pc}$.  Using the $J$-band offset of 1.56, we find $d = 430^{+340}_{-190}\,\mathrm{pc}$.

\section{The space density of CVs}
\label{sec:rho}
Here, we give an estimate of the space density of CVs based on our sample (Section~\ref{sec:oneovervmax}), and also consider, in Section~\ref{sec:limits}, how large a population of systems with given (low) $L_X$ could have escaped detection in the NEP survey.

DNe typically have softer X-ray spectra in outburst than in quiescence, although a wide range in X-ray behaviour has been observed in different systems (e.g. \citealt{bible}).  The RASS was conducted over a period of about 6 months, and during this time the NEP area was observed many times.  This means that, while the duration of the survey was short compared to the outburst recurrence time-scale of faint DNe, high duty cycle, frequently outbursting CV, such as EX Dra, were probably observed both in outburst and quiescence.  We will not attempt to account for this complication.  

\begin{table*}
 \centering
  \caption{Orbital periods, X-ray fluxes, distances, and luminosities of the five CVs in the NEP survey area, together with the contribution each systems makes (if errors are ignored) to the mid-plane space density.  Note that, although IX Dra is in the area covered by the survey, it is not included in the NEP sample.}
  \label{tab:cvs}
  \begin{tabular}{@{}lllllllll@{}}
  \hline
Object               & $P_{orb}/\mathrm{h}$ & $F_X/\mathrm{erg\,cm^{-2}\,s^{-1}}$ & $d/\mathrm{pc}$ & $L_X/\mathrm{erg\,s^{-1}}$ & $(1/V_{max})/\mathrm{pc^{-3}}$\\                       
 \hline
EX Dra          & 5.0385& $(6.4 \pm 0.7) \times10^{-14}$ & $260^{+140}_{-90}$  & $5\times10^{29}$ & $8.8 \times 10^{-6}$ \\[0.1cm]
RX J1831.7+6511 & 4.01  & $(1.7 \pm 0.2) \times10^{-13}$ & $980^{+630}_{-380}$ & $2\times10^{31}$ & $1.4 \times 10^{-6}$ \\[0.1cm]
SDSS J1730      & 1.837 & $(3.4 \pm 0.3) \times10^{-13}$ & $440^{+280}_{-170}$ & $8\times10^{30}$ & $3.1 \times 10^{-7}$ \\[0.1cm]
RX J1715.6+6856 & 1.64  & $(8.5 \pm 1.7) \times10^{-14}$ & $400^{+400}_{-200}$ & $2\times10^{30}$ & $1.3 \times 10^{-6}$ \\[0.3cm]
IX Dra          & 1.595 & $(2.2 \pm 0.5) \times10^{-14}$ & $430^{+340}_{-190}$ & $5\times10^{29}$ & $5.4 \times 10^{-6}$ \\
 \hline
 \end{tabular}
\end{table*}

We assume that the vertical density profile of CVs is exponential
\begin{equation}
\rho(z)=\rho_0 \mathrm{e}^{-|z|/h},
\label{eq:z}
\end{equation}
with $z$ the perpendicular distance from the Galactic plane (i.e. $z=d \sin b$, where $b$ is Galactic latitude).  The local space density is defined as $\rho_0=\rho(0)$, the mid-plane value of $\rho$.  We will ignore any radial dependence of $\rho$ in Section~\ref{sec:oneovervmax} (it certainly has a negligible effect on the results presented there), but include it in Section~\ref{sec:limits} (simply because it is not too computationally difficult).

Widely used empirical values of the Galactic scale height of CVs, $h$, may be found in \cite{Patterson84} and \cite{Duerbeck84}.  However, since all observational CV samples contain a serious $z$-dependent bias, we will use theoretical $h$ values instead.  Also, rather than assuming a single scale height for all CVs, we will follow \cite{PretoriusKniggeKolb07} in setting $h=120$, $260$, and $450\,\mathrm{pc}$ for long-period systems, short-period systems, and period bouncers, respectively\footnote{This is clearly still a very simplified picture, but it is an improvement over assuming a single small scale height for the entire CV population.}.

CVs are expected to have X-ray emission that can be described as thermal bremsstrahlung with $kT$ roughly between 5 and 20~keV \citep{PattersonRaymond85}; observations indicate a larger variety of X-ray spectral shapes (e.g. \citealt{VrtilekSilberRaymond94}; \citealt{Richman96}; \citealt{NaylorBathCharles88}).  The range in $F_X$ resulting from varying $kT$ from 5 to 20~keV for a thermal bremsstrahlung spectrum is about as large as the observational error (which is an insignificant contribution to our uncertainty in $\rho_0$).  We will therefore simply assume a $kT=10\,\mathrm{keV}$ thermal bremsstrahlung spectrum for all CVs.  We follow \cite{Henry06} in quoting X-ray fluxes and luminosities in the 0.5--2.0~keV band, calculated from \textit{ROSAT} PSPC count rates in the 0.1--2.4~keV band.

\subsection{The $1/V_{max}$ method}
\label{sec:oneovervmax}
The space density is found by counting the CVs inside the volume observed, while accounting for both the fact that our sample is flux- rather than volume-limited, and the variation of space density with $z$.  The method involves essentially allowing the `volume limit' to vary according to the luminosities of the systems in our sample, while the dependence on $z$ is taken care of by defining a `generalised' volume
\begin{equation}
V_j=\Omega \frac{h^3}{|\sin b|^3}\left[2-\left(x_j^2+2x_j+2\right)\mathrm{e}^{-x_j}\right]
\label{eq:V}
\end{equation}
(e.g. \citealt{Schmidt68}; \citealt{Felten76}; \citealt{StobieIshidaPeacock89}; \citealt{TinneyReidMould93}), where $\Omega$ is the solid angle covered by the survey and $x_j=d_j |\sin b|/h$, with $d_j$ the maximum distance at which a CV included in the sample could have been detected (given its $L_X$ and the survey flux limit).  Equation~\ref{eq:V} assumes that $\rho$ is a function only of $z$, and that this dependence is given by equation~\ref{eq:z}.  The index $j$ represents the CVs in our sample.  Because both $b$ and the flux limit are variable over $\Omega$, we compute each $V_j$ as a sum over solid angles $\delta\Omega$ small enough so that $b$ and the flux limit (and thus $d_j$) can be assumed to be constant over each $\delta\Omega$.  This is done by dividing the NEP area into a $36 \times 36$ pixel grid.  $\rho_0$ is then the sum of the space densities represented by each CV, i.e.
$$\rho_0=\sum_j 1/V_j.$$
The $1/V_{max}$ values for the 5 CVs, obtained by using the best distance and $F_X$ estimates, are given in Table~\ref{tab:cvs}, together with other relevant parameters.

We neglect extinction in finding the maximum distance at which a source could be detected from its estimated distance (for which extinction was included) and observed $F_X$.  This is reasonable because extinction is low in the NEP area (the absorbed $F_X$ is at least 0.78 times the unabsorbed flux for all positions in the survey volume), and because, for the distances of most of our sources, the column densities are already at close to the total Galactic values.  In other words, at distances beyond our sources, extinction is approximately constant, so that the maximum volumes we obtain are unaffected if extinction is ignored.

Both distance errors and the small sample size contribute significantly to the overall uncertainty on $\rho_0$, although the latter effect dominates. In order to account fully for all sources of statistical error (including the uncertainty in distances, and also the error on $F_X$), we determine the probability distribution of $\rho_0$ via a Monte Carlo simulation. Thus we create a large number of mock CV samples with properties designed to fairly sample the full parameter space permitted by the data. We then compute an estimate of $\rho_0$ for each mock sample, thus gradually generating a probability distribution for $\rho_0$.

The key step in this Monte Carlo simulation is the generation of appropriate mock samples. This is done by treating each CV in the real sample as a representative of a larger population of similar CVs, and allowing each of these populations to contribute to each mock sample. In creating a particular mock sample, we therefore consider each CV in the real sample in turn, and generate a mock counterpart to it by drawing the X-ray flux and distance from the appropriate probability distributions.  This takes care of the observational uncertainties in $F_X$, as well as the uncertainties in distances.  However, there is also Poisson uncertainty associated with the small sample size, which would be ignored if each mock sample contained exactly the same number of CVs as the real sample.  We deal with this by assigning a weighting factor to each counterpart CV in each mock sample. The weighting factor needs to be drawn independently for each counterpart CV from the probability distribution of the expected number of sources belonging to this population in a NEP-like survey.  This probability distribution can be constructed from Bayes' theorem. If the expected number of sources in a NEP-like survey is $\mu$ for a particular population\footnote{Note that $\mu$ does not need to be an integer.}, the actual number observed in a particular survey will be $N_{obs}$, where this is drawn from the Poisson distribution $P(N_{obs}|\mu)$. Then Bayes' theorem gives
$$P(\mu|N_{obs}) \propto P(N_{obs}|\mu) P(\mu),$$
where $P(\mu)$ is the prior on $\mu$. Note that, in practice, $N_{obs} \equiv 1$ by construction for each population represented in the real sample (since each CV is taken to represent a population of similar CVs). Also, since $\mu$ is a scale factor for the space density of each population, the appropriate uninformative prior to use is $P(\mu) = 1/\mu$ (e.g. \citealt{Jeffreys1961})\footnote{We carried out several tests to check that this technique produces reliable error estimates.  In the tests, we take a space density, luminosity function, and flux limit as input, and use these to simulate observed samples.  The $\rho_0$ distribution implied by each sample is then calculated in the same way as above.  We found that for $\simeq$68\% of these $\rho_0$ distributions the true input $\rho_0$ is contained in any particular 1-$\sigma$ interval.  This holds when the expected number of detected sources is as small as 4, but also for much larger samples, when the distributions become narrower.  We ignored Galactic structure in these tests (i.e. $\rho$ was taken to be constant), but experimented with several different (simple) luminosity functions, as well as with including errors in luminosity.  We also ran some simulations to verify that a uniform prior is not appropriate in this calculation.  These numerical tests indicate that the method we use is reliable, and that we have adopted a suitable prior.
}.

\begin{figure}
 \includegraphics[width=84mm]{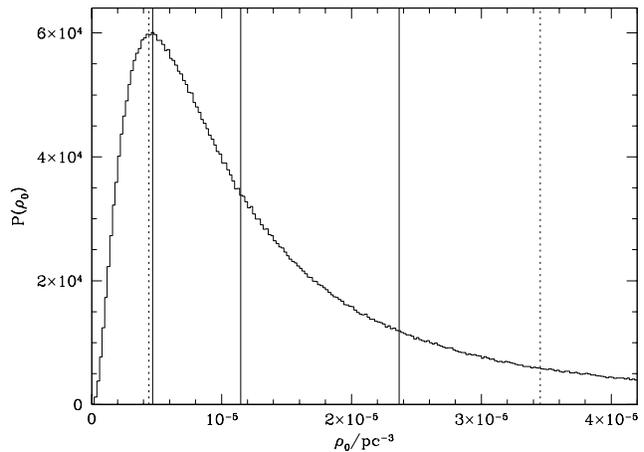} 
  \caption {The $\rho_0$ distribution resulting from our simulation.  This sample includes RX J1715.6+6856, SDSS J1730, RX J1831.7+6511, and EX Dra; we used scale heights of $260\,\mathrm{pc}$ and $120\,\mathrm{pc}$ for short- and long-period systems, respectively.  Solid lines mark the mode, median, and mean at $4.7 \times 10^{-6}$, $1.1 \times 10^{-5}$, and $2.4 \times 10^{-5}\,\mathrm{pc^{-3}}$, and dotted lines show a 1-$\sigma$ interval from $4.4 \times 10^{-6}$ to $3.5 \times 10^{-5}\,\mathrm{pc^{-3}}$.
}
 \label{fig:rho_pdf}
\end{figure}

To summarise, we draw a random distance and $F_X$ from the appropriate probability distribution functions (assuming that the error in $F_X$ is Gaussian) for each of the systems in the sample, compute $V_j$, generate $\mu_j$, and obtain $\rho_0$ by summing $\mu_j/V_j$ over $j$.  Repeating this a large number of times produces a probability distribution for $\rho_0$.  Fig.~\ref{fig:rho_pdf} shows the $\rho_0$ histogram obtained from this calculation.  The mode, median, and mean of this distribution are $4.7 \times 10^{-6}$, $1.1 \times 10^{-5}$, and $2.4 \times 10^{-5}\,\mathrm{pc^{-3}}$, and are marked by solid lines.  The dotted lines show the symmetric 1-$\sigma$ confidence interval around the median, which is $4.4 \times 10^{-6}$ to $3.5 \times 10^{-5}\,\mathrm{pc^{-3}}$.

A reason for some concern is that IX Dra was only just too faint to be included in the NEP sample (the fact that it might be a period bouncer adds to this problem).  More importantly, a single system, EX Dra, dominates the space density estimate.  Our choice of scale heights is also non-standard.  To investigate the sensitivity of the result to these factors, we list in Table~\ref{tab:rhos} the space densities that would be obtained if we (i) used a single scale height of $120\,\mathrm{pc}$, (ii) included IX Dra, and (iii) excluded EX Dra.  Note that as different as the best estimates (we give the median of each distribution) are, they are all consistent with each other.  Table~\ref{tab:rhos} also gives the space densities for short- and long-period systems separately.

\begin{table}
 \centering
  \caption{Systems are identified as 1---EX Dra, 2---RX J1831.7+6511, 3---SDSS J1730, 4---RX J1715.6+6856, and 5---IX Dra in the first column.  $h_1$ and $h_2$ are the scale heights, in parsecs, assumed for short- and long-period systems, respectively.  The values given for $\rho_0$ are the medians of the $\rho_0$ distributions, and the errors represent symmetric 1-$\sigma$ confidence intervals, around the median.}
  \label{tab:rhos}
  \begin{tabular}{@{}ll@{}}
  \hline
Sample and scale height & $\rho_0/\mathrm{pc^{-3}}$ \\
 \hline
systems 1, 2, 3, 4; $h_1=260$, $h_2=120$    & $1.1^{+2.3}_{-0.7}\times10^{-5}$  \\[0.1cm]
systems 1, 2, 3, 4; $h_1=h_2=120$           & $1.5^{+2.5}_{-0.8}\times10^{-5}$  \\[0.1cm]
systems 1, 2, 3, 4, 5; $h_1=260$, $h_2=120$ & $2.1^{+4.2}_{-1.3}\times10^{-5}$  \\[0.1cm]
systems 2, 3, 4; $h_1=260$, $h_2=120$       & $3.1^{+5.5}_{-1.9}\times10^{-6}$  \\[0.1cm]
systems 1, 2; $h_2=120$ (long-period)       & $7.7^{+18}_{-5.1}\times10^{-6}$   \\[0.1cm]
systems 3, 4; $h_2=260$ (short-period)      & $1.4^{+5.2}_{-1.0}\times10^{-6}$  \\
 \hline
 \end{tabular}
\end{table}

\subsection{Upper limits on the space density of an undetected population}
\label{sec:limits}
An unavoidable assumption of the $1/V_{max}$ method is that the detected objects are representative of the luminosity function of the true underlying CV population.  No evidence of even a large population of CVs is expected to show up in a flux-limited survey, if such a population is sufficiently faint.

The faintest measured values of $L_X$ for CVs are a few times $10^{29}\,\mathrm{erg\,s^{-1}s}$ (e.g. \citealt{VerbuntBunkRitter97}; \citealt{WheatleyBurleighWatson00}; \citealt{SchwopeStaudeKoester07})\footnote{\cite{VerbuntBunkRitter97} list $L_X=3.16 \times 10^{28}\,\mathrm{erg\,s^{-1}}$ for AY Lyr, but this resulted from a very low distance estimate of $57\,\mathrm{pc}$ \citep{SproatsHowellMason96}.  The distance is based on a measurement of $K=13.08$, taken during outburst.  2MASS photometry indicates $K_S>15.6$ in quiescence, and using the \cite{Knigge06} sequence (with no offset), we find $d>277\,\mathrm{pc}$, implying that the system is not particularly faint in X-rays.}.  However, we do not really know how faint CVs can be: the faintest expected secular luminosities can be estimated from the gravitational radiation $\dot{M}$, but, since faint CVs are DNe, they spend most of their time at fainter luminosities.

\cite{RiazGizisHarvin06} find $10^{27}\,\mathrm{erg\,s^{-1}} < L_X < 10^{33}\,\mathrm{erg\,s^{-1}}$ for a large sample of nearby dwarfs with spectral types between K5 and M6; brown dwarfs have $L_X < 10^{29}\,\mathrm{erg\,s^{-1}}$ (often several orders of magnitude less than $10^{29}\,\mathrm{erg\,s^{-1}}$; e.g. \citealt{StelzerMicelaFlaccomio06}).  CV secondaries have shorter spin periods than expected for the objects in these studies, but, even so, we cannot be confident that the secondary star alone guarantees $L_X > 10^{29}\,\mathrm{erg\,s^{-1}}$.  This is especially true in the case of period bouncers.

We use another Monte Carlo simulation to find the upper limit on the space density of a hidden population, assuming that all members of this population have the same (low) $L_X$.
The Galaxy is modelled as an axisymmetric disc, with exponential vertical and radial ($r$) density profiles.  The galactic centre distance is taken as $7\,620\,\mathrm{pc}$ \citep{EisenhauerGenzelAlexander05}, and the small offset of the Sun from the galactic plane is neglected.  At least part of a hidden population may consist of pre-period bounce systems, but we conservatively assume a single scale height of $450$~pc (appropriate to period bouncers).  The radial scale length is taken as 3\,000~pc.
The local space density is then $\rho_0$, where $\rho = \rho_0 \exp (-|z|/450\,\mathrm{pc})$, for $r$ near $7\,620\,\mathrm{pc}$.

Because our upper limits will be systematically low if extinction is neglected, we include it here, using the model of \cite{AmoresLepine05} to find $N_H$ for every simulated system.

\begin{figure}
 \includegraphics[width=84mm]{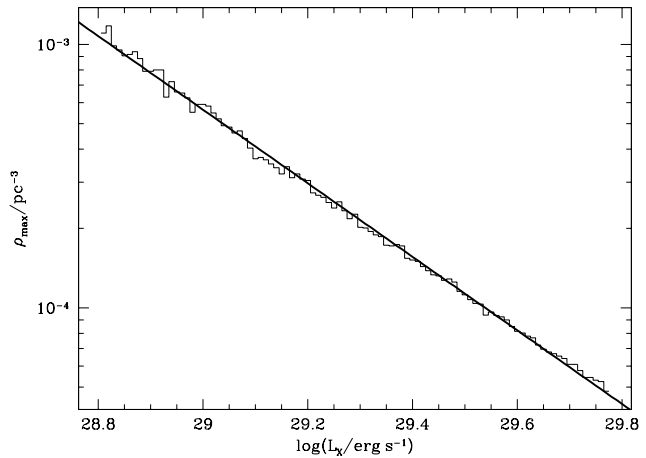} 
  \caption {The upper limit on the mid-plane space density as a function of X-ray luminosity for an unseen population of CVs.  The data from the simulation are shown as a fine histogram, and a fit is over-plotted as a bold curve.
}
 \label{fig:limit}
\end{figure}

We then find the value of $\rho_0$ for which the predicted number of faint systems detected in the NEP survey is 3 (detecting 0 such systems is then a 2-$\sigma$ result).
Fig.~\ref{fig:limit} shows the maximum allowed $\rho_0$ as a function of $L_X$ for CVs that make up the hidden population.  Results from the simulation are plotted as a fine histogram, while the bold curve is a fit to the data.  The fit is given by 
$$\rho_{max}= 5.7\times 10^{-4} (L_x/10^{29}\,\mathrm{erg\,s^{-1}})^{-1.39}\,\mathrm{pc^{-3}}.$$
Thus for $L_X =2 \times 10^{29}\,\mathrm{erg\,s^{-1}}$, we have $\rho_0 < 2 \times 10^{-4}\,\mathrm{pc^{-3}}$.  Note that here $\rho_0$ refers only to the hypothesized undetected population, and does not include the contribution from the brighter, detected systems.  

\section{Discussion}
The $\rho_0$ measurement from the NEP survey is consistent with several previous observational estimates, e.g. $\sim 10^{-5}\mathrm{pc}^{-3}$ \citep{Patterson98} and $\sim 3 \times 10^{-6}\mathrm{pc}^{-3}$ \citep{brian01}.  A particularly relevant study to compare to is \cite{HertzBailynGrindlay90}, since it was also based on a sample of 4 X-ray selected CVs.  Hertz et al. find $\rho_0 \simeq (2-3) \times 10^{-5}\mathrm{pc}^{-3}$ (where the uncertainty includes only the possible incompleteness of the sample), again consistent with the estimate presented here.

The most important concern with our measurement is that the NEP sample is so small that it may not fairly represent the underlying CV population.  This is apparent from the fact that we estimate a higher space density for long- than for short-period systems.  The reason is that our estimate is dominated by the long-period system EX Dra, which has an unusually low $L_X$ for its orbital period, probably because of its very high inclination.  Note, however, that if the X-ray luminosity of EX Dra had been higher, that would only reduce $\rho_0$.

Theoretical predictions of the CV space density are typically higher than most observational values.  A space density of $(0.5 - 2) \times 10^{-4}\,\mathrm{pc^{-3}}$ results from the CV birth rate of \cite{deKool92}, together with the assumptions that the age of the Galaxy is 10 Gy, and that the lifetime of a CV is longer than that \citep{bible}.  When the model population of \cite{Kolb93} is normalised to the formation rate of single white dwarfs, the predicted mid-plane CV space density is $1.8 \times 10^{-4}\,\mathrm{pc^{-3}}$.  We cannot rule out these values, but our data do show that if $\rho_0$ is really as high as $2 \times 10^{-4}\,\mathrm{pc^{-3}}$, then the dominant CV population must have $L_X \la 2 \times 10^{29}\,\mathrm{erg\,s^{-1}}$.  It is suspected that the mass transfer rates of CVs fluctuate around the secular mean on time-scales much longer than the total observational base line (see e.g. \citealt{HameuryKingLasota89}; \citealt{WuWickramasingheWarner95}; \citealt{KingFrankKolb95}; \citealt{McCormickFrank98}; \citealt{BuningRitter04}; \citealt{RitterZhangKolb00}).  Because the time-scale of these $\dot{M}$ cycles is not accessible to observations, we can only note that surveys useful for finding CVs typically have no sensitivity to objects that are not currently accreting.

The fact that stellar populations of different ages do not have the same Galactic distribution has to be kept in mind when interpreting (and, indeed, measuring) space densities.  While the situation is clearly not really as simple as three distinct sub-populations with three scale heights, long-period CVs make up a significantly younger population than, say, period bouncers.  One should therefore clearly not expect the ratio of the mid-plane space densities of these sub-populations to be the same as the ratio of their total Galactic numbers.  This may be a trivial point, but it has been completely ignored in the literature.  If, e.g., we assume that 70\% of all CVs in the Galaxy are period bouncers, and only 1\% are long-period systems, and if the space densities of these two sub-populations are given by $\rho_{pb}(z)=\rho_{pb}(0) \exp(-|z|/450\,\mathrm{pc})$  and $\rho_{lp}(z)=\rho_{lp}(0) \exp(-|z|/120\,\mathrm{pc})$, respectively, then (from integrating the vertical density profiles and assuming the radial and azimuthal dependences are the same in both cases) we have $\rho_{pb}(0)=70(120/450)\rho_{lp}(0)\simeq 20\rho_{lp}(0)$.  We can also use this argument to place a tentative limit on the ratio of the total numbers of long-period CVs and periods bouncers.  If we assume $\rho_{pb}(0)<2 \times 10^{-4}\,\mathrm{pc^{-3}}$ and $\rho_{lp}(0)>2.0 \times 10^{-6}\,\mathrm{pc^{-3}}$, then this ratio must be less than about 30.

There are several ongoing surveys that will produce large new samples of CVs, e.g. the Hamburg Quasar Survey, the Sloan Digital Sky Survey, and the INT Photometric H$\alpha$ Survey of the Northern Galactic Plane (e.g. \citealt{GansickeHagenEngels02}; \citealt{sdsscvs1}; \citealt{WithamKniggeGansicke06}).  Although these samples will all be subject to more complicated selection effects than the NEP sample, they will be much larger, which will remove the main shortcoming of the present study (although it should be kept in mind that space density estimates based on larger samples can still be dominated by 1 or 2 systems; e.g. \citealt{SchwopeBrunnerBuckley02}).  The Palomar Green Survey CV sample (e.g. \citealt{Ringwald93}) is already available, and is also suitable for a rigorous measurement of $\rho_0$, since it is complete and has well-defined selection criteria.

\section{Conclusions}
Using the complete X-ray flux-limited CV sample detected in the \emph{ROSAT} NEP survey, we have estimated the space density of CVs bright enough to be represented in the survey, and placed limits on the size of a fainter population that may have remained undetected.

We have presented observations of two newly discovered CVs, namely RX J1715.6+6856 and RX J1831.7+6511, and measured orbital periods of 1.64~h and 4.01~h respectively for these systems.  RX J1715.6+6856 and RX J1831.7+6511 are probably both DNe.  This completes the identification of CVs in the \emph{ROSAT} NEP survey.

Distances were estimated for the 4 systems in the NEP sample, as well as for IX Dra, which is in the area covered by the survey, but below the survey flux limit.

With the assumption that the 4 CVs included in the NEP survey are representative of the intrinsic CV population, the space density of CVs is $1.1^{+2.3}_{-0.7} \times 10^{-5}\,\mathrm{pc^{-3}}$.  Space densities as high as the theoretically predicted $\simeq 2 \times 10^{-4}\,\mathrm{pc^{-3}}$ are excluded by our data, unless the dominant CV population is fainter than $\simeq 2 \times 10^{29}\,\mathrm{erg\,s^{-1}}$ in X-rays.

We also emphasise the fact that long-period CVs, short-period CVs, and period bouncers have different Galactic scale heights; this needs to be accounted for when mid-plane space densities are translated to total Galactic number of systems.

\section*{Acknowledgements}
MLP acknowledges financial support from the South African National Research Foundation and the University of Southampton.  We thank Giusi Micela for communicating results of the follow-up of stellar NEP sources in advance of publication, and Brian Warner for commenting on the manuscript.  Our thanks also to John Thorstensen for giving us the orbital period of SDSS J173008.38+624754.7, and to Joe Patterson for a discussion of IX Dra.

\bsp

\label{lastpage}

\bibliographystyle{mn2e}

\end{document}